\documentclass[11pt]{article}

\usepackage[margin=1in]{geometry}
\usepackage{booktabs}
\usepackage{graphicx}
\graphicspath{{}}
\usepackage{amsmath,amssymb}
\usepackage{microtype}
\usepackage{hyperref}
\usepackage{cleveref}
\usepackage{algorithm}
\usepackage{algpseudocode}
\usepackage{tensors}
\usepackage{xcolor}
\usepackage{xspace}
\usepackage{authblk}
\usepackage{float}
\usepackage{placeins}
\usepackage{tikz}
\usetikzlibrary{arrows.meta, calc, decorations.pathreplacing, patterns}

\newcommand{\tsvdmi}{t-SVDM-I\xspace}
\newcommand{\tsvdmii}{t-SVDM-II\xspace}

\begin{document}

\title{High-Performance $\starM{}$ SVD for Big Data Compression}

\author[1]{Md Taufique Hussain}
\author[1]{Grey Ballard}
\author[1]{Aditya Devarakonda}
\author[2]{Srinivas Eswar}
\author[3]{Naman Pesricha}
\author[2]{Vishwas Rao}

\affil[1]{Wake Forest University, Winston-Salem, NC, USA}
\affil[2]{Argonne National Laboratory, Lemont, IL, USA}
\affil[3]{Indian Institute of Science, Bengaluru, India}

\date{}

\maketitle

\begin{abstract}
In the era of big data, effectively compressing large datasets while performing complex mathematical operations is crucial.
Tensor-based decomposition methods have shown superior compression capabilities with minimal loss of accuracy compared to traditional matrix methods.
Under the $\starM{}$ tensor framework, tensors can be decomposed in a matrix-mimetic way, including using the $\starM{}$ SVD.
This tensor SVD has optimality guarantees and has shown exceptional performance on specific types of data, but software implementations have been mostly limited to productivity-oriented languages.
In this work, we present our development of a shared-memory parallel, high-performance solution designed to efficiently implement the underlying algorithms.
This software will enable optimal compression of extensive scientific datasets, paving the way for enhanced data analysis and insights.
\end{abstract}



\section{Introduction}\label{sec:intro}

Truncated tensor-based singular value decompositions based on the star-M ($\starM{}$) tensor-tensor product mimic the matrix SVD \cite{kilmer2021tensor}.
An input tensor can be decomposed into the product of three factor tensors with properties of orthogonality and structure that are analogous to the matrix case.
Furthermore, this decomposition can be truncated for the purpose of data compression, for example, in a systematic way to determine the optimal representation given either an error tolerance or specified rank parameters.
This is possible due to the existence of theoretical guarantees established for the $\starM{}$ tensor algebra that generalize the Eckart-Young theorem \cite{eckart1936approximation} for low-rank matrix approximation.

As described in more detail in \Cref{sec:background}, given 3-way input tensors, the $\starM{}$ product is parametrized by an invertible matrix $\Mx{M}$ and defined by a transformation of the mode-3 fibers of the inputs by $\Mx{M}$, independent matrix multiplications of the mode-3 slices, and a back transformation of the output mode-3 fibers by $\Mx{M}^{-1}$.
The $\starM{}$ truncated SVD (t-SVD) is thus based on slice-wise matrix SVDs after transformation by the matrix $\Mx{M}$, and it is particularly advantageous when the transformation yields many slices to be either low-rank or insignificant relative to other slices.
The $\starM{}$ t-SVD is not agnostic with respect to mode ordering, and the choice of $\Mx{M}$ can have a substantial effect on the compressibility of a data set.
Implementations of $\starM{}$ algebra and the $\starM{}$ t-SVD exist in MATLAB and Python---see \cite{Lu18,newman2025optimal,KVX21}, for example---but no high performance implementations have been developed.
High performance, parallel software exists for many other tensor formats including CANDECOMP/PARAFAC (CP) \cite{HL+21,EH+21,WKP25}, Tucker \cite{BKK20,KU16}, and Tensor Train \cite{ABB22,RTB22}.
The goal of this work is to design an efficient and scalable shared-memory parallel implementation of the fundamental algorithms for computing $\starM{}$ t-SVD for general multiway data, enabling its use for datasets that were otherwise too large to handle using existing techniques.

The core of the $\starM{}$ t-SVD algorithm is a ``batch'' of independent matrix SVDs, all of the same dimension, which is ripe for parallelization.
In fact, implementations of a batched SVD are emerging in standard libraries like MKL \cite[Version 2024.1]{MKL} and MAGMA \cite[Version 2.10.0]{dghklty14}.
However, the (back-)transformations of the slices necessarily happens in the orthogonal direction, across slices, enabling the decoupling.
We show in \Cref{sec:implementation} that the transformations can be parallelized via efficient tensor-times-matrix operations that respect the data layout of the tensor in memory and exploits the interface to the batched BLAS \cite{AC+21}.
Two primary algorithms exist for t-SVD: the first specifies a constant rank for the truncation of all slices, and the second specifies an error tolerance that guides the truncation to various ranks across slices.
We show how to parallelize the first algorithm using the expert interface to LAPACK, and we explore computation-memory tradeoffs in the design of the second algorithm, which requires analysis of the computed slice-wise singular values to determine ranks.

In \Cref{sec:application}, we consider data sets coming from the domains of climate reanalyses, computational fluid dynamics, and X-ray crystallography.
We describe how mode ordering and choice of transformation matrix can affect both performance of the parallel implementations and the quality of the compressed representation. 
In the case of the climate data, we show the benefits of using tensor-based compression over the standard matrix-based approach used in that field.
Our parallel design decisions enable efficient parallel scaling, and we show that on a platform with 64 cores, these datasets can be compressed up to $42\times$, $40\times$, and $30\times$ faster than a single core, respectively.
Our implementation (which will be made publicly available) enables users to observe speedups of up to $1000\times$ over existing MATLAB code on a similar machine.
Thus, this implementation allows domain scientists to analyze larger data sets in less time, enabling the $\starM{}$ t-SVD to be applied as broadly as other tensor decomposition techniques.

\section{Background}\label{sec:background}
This section reviews the mathematical and algorithmic background relevant to our work.
We first define notation for tensor and matrix-tensor operations, and define the \starM{}-product (\Cref{subsec:notation}).
We then introduce the \starM{} t-SVD together with its full (\tsvdmi) and adaptively truncated (\tsvdmii) variants, whose parallelization and evaluation is the main focus of our work (\Cref{subsec:tsvd}).
Finally, we summarize the Eckart-Young-style optimality results that distinguish the \starM{} t-SVD from competing tensor decompositions (\Cref{subsec:optimality}) and motivate the need for a shared-memory parallel implementation suited to large scientific datasets (\Cref{subsec:motivation}).

\begin{figure}[H]
    \centering
%

\begin{tikzpicture}[
  font=\small,
  >=Stealth,
]

\def\sliceH{2.0}    
\def\fullR{1.4}     
\def\dx{0.17}       
\def\dy{0.14}       
\def\Nlayers{5}     
\pgfmathtruncatemacro{\backLayer}{\Nlayers-1}
\def\depthArrowPad{0.14} 
\def\panelGap{0.45cm}    

\def\rDim{1.2}    

\def\rankA{1.05}   
\def\rankB{0.32}   
\def\rankC{1.30}   
\def\rankD{0.65}   
\def\rankE{1.15}   

\colorlet{Ucol}{blue!18}
\colorlet{Ubord}{blue!55!black}
\colorlet{Gcol}{orange!18}
\colorlet{Gbord}{orange!60!black}
\colorlet{wastecol}{gray!12}
\colorlet{wastebord}{gray!45}

\begin{scope}

  \foreach \i in {4,3,2,1,0} {
    \fill[Ucol]  ({\i*\dx}, {\i*\dy}) rectangle ++({\rDim}, {\sliceH});
    \draw[Ubord] ({\i*\dx}, {\i*\dy}) rectangle ++({\rDim}, {\sliceH});
  }
  \draw[<->, thin] (-0.25, 0) -- (-0.25, \sliceH)
    node[midway, left, font=\footnotesize] {$m$};
  \draw[<->, thin] (0, -0.28) -- (\rDim, -0.28)
    node[midway, below, font=\footnotesize] {$r$};
  \draw[<->, gray, thin] ({\backLayer*\dx + \rDim + 0.08 + \depthArrowPad}, {\backLayer*\dy + \sliceH + 0.05 + \depthArrowPad})
    -- ({\rDim + 0.08 - \depthArrowPad}, {\sliceH + 0.05 - \depthArrowPad});
  \node[gray, font=\footnotesize, right] at ({\backLayer*\dx + \rDim + 0.1 + \depthArrowPad}, {\backLayer*\dy + \sliceH + 0.05 + \depthArrowPad})
    {$n$};
  \node[below, align=center, font=\footnotesize] at ({\rDim*0.5 + 2*\dx}, -0.55)
    {$\hat{\Tn{U}}\in\mathbb{R}^{m\times r\times n}$};

  \begin{scope}[xshift=2.75cm]
    \foreach \i in {4,3,2,1,0} {
      \fill[Gcol]  ({\i*\dx}, {\i*\dy}) rectangle ++({\fullR}, {\rDim});
      \draw[Gbord] ({\i*\dx}, {\i*\dy}) rectangle ++({\fullR}, {\rDim});
    }
    \draw[<->, thin] (-0.25, 0) -- (-0.25, {\rDim})
      node[midway, left, font=\footnotesize] {$r$};
    \draw[<->, thin] (0, -0.28) -- (\fullR, -0.28)
      node[midway, below, font=\footnotesize] {$p$};
    \draw[<->, gray, thin] ({\backLayer*\dx + \fullR + 0.08 + \depthArrowPad}, {\backLayer*\dy + \rDim + 0.05 + \depthArrowPad})
      -- ({\fullR + 0.08 - \depthArrowPad}, {\rDim + 0.05 - \depthArrowPad});
    \node[gray, font=\footnotesize, right] at ({\backLayer*\dx + \fullR + 0.1 + \depthArrowPad}, {\backLayer*\dy + \rDim + 0.05 + \depthArrowPad})
      {$n$};
    \node[below, align=center, font=\footnotesize] at ({\fullR*0.5 + 2*\dx}, -0.55)
      {$\hat{\Tn{G}}\in\mathbb{R}^{r\times p\times n}$};
  \end{scope}

  \node[above, font=\normalsize\bfseries] at ({(2.3 + \fullR)*0.5 + 2*\dx}, {4*\dy + \sliceH + 0.75})
    {t-SVDM-I};
  \node[above, font=\footnotesize] at ({(2.3 + \fullR)*0.5 + 2*\dx}, {4*\dy + \sliceH + 0.42})
    {uniform rank $r$ per slice};

\end{scope}

\begin{scope}[xshift={6.2cm + \panelGap}]

  \foreach \i/\rk in {4/\rankE, 3/\rankD, 2/\rankC, 1/\rankB, 0/\rankA} {
    \fill[wastecol] ({\i*\dx}, {\i*\dy}) rectangle ++({\rankC}, {\sliceH});
    \draw[wastebord, dashed, thin] ({\i*\dx}, {\i*\dy}) rectangle ++({\rankC}, {\sliceH});
    \fill[Ucol]  ({\i*\dx}, {\i*\dy}) rectangle ++({\rk}, {\sliceH});
    \draw[Ubord] ({\i*\dx}, {\i*\dy}) rectangle ++({\rk}, {\sliceH});
  }

  \draw[<->, thin] (-0.25, 0) -- (-0.25, \sliceH)
    node[midway, left, font=\footnotesize] {$m$};
  \draw[<->, thin] (0, -0.28) -- (\rankA, -0.28)
    node[midway, below, font=\footnotesize] {$\rho_1$};
  \draw[<->, thin] ({4*\dx}, {4*\dy + \sliceH + 0.12}) -- ({4*\dx + \rankE}, {4*\dy + \sliceH + 0.12})
    node[midway, above, font=\footnotesize] {$\rho_n$};
  \draw[<->, gray, thin] ({\backLayer*\dx + \rankE + 0.08 + \depthArrowPad}, {\backLayer*\dy + \sliceH + 0.05 + \depthArrowPad})
    -- ({\rankE + 0.08 - \depthArrowPad}, {\sliceH + 0.05 - \depthArrowPad});
  \node[gray, font=\footnotesize, right] at ({\backLayer*\dx + \rankE + 0.1 + \depthArrowPad}, {\backLayer*\dy + \sliceH + 0.05 + \depthArrowPad})
    {$n$};
  \node[below, align=center, font=\footnotesize] at ({\rankA*0.5 + 2*\dx}, -0.55)
    {$\hat{\Tn{U}}$ (variable rank)};

  \begin{scope}[xshift=2.75cm]
    \foreach \i/\rk in {4/\rankE, 3/\rankD, 2/\rankC, 1/\rankB, 0/\rankA} {
      \fill[wastecol] ({\i*\dx}, {\i*\dy}) rectangle ++({\fullR}, {\rankC});
      \draw[wastebord, dashed, thin] ({\i*\dx}, {\i*\dy}) rectangle ++({\fullR}, {\rankC});
      \fill[Gcol]  ({\i*\dx}, {\i*\dy + \rankC - \rk}) rectangle ++({\fullR}, {\rk});
      \draw[Gbord] ({\i*\dx}, {\i*\dy + \rankC - \rk}) rectangle ++({\fullR}, {\rk});
    }
    \draw[<->, thin] (-0.25, {\rankC - \rankA}) -- (-0.25, {\rankC})
      node[midway, left, font=\footnotesize] {$\rho_1$};
    \draw[<->, thin] (0, -0.28) -- (\fullR, -0.28)
      node[midway, below, font=\footnotesize] {$p$};
    \draw[<->, thin] ({4*\dx + \fullR + 0.12}, {4*\dy + \rankC - \rankE}) -- ({4*\dx + \fullR + 0.12}, {4*\dy + \rankC})
      node[midway, right, font=\footnotesize] {$\rho_n$};
    \draw[<->, gray, thin] ({\backLayer*\dx + \fullR + 0.08 + \depthArrowPad}, {\backLayer*\dy + \rankC + 0.05 + \depthArrowPad})
      -- ({\fullR + 0.08 - \depthArrowPad}, {\rankC + 0.05 - \depthArrowPad});
    \node[gray, font=\footnotesize, right] at ({\backLayer*\dx + \fullR + 0.1 + \depthArrowPad}, {\backLayer*\dy + \rankE + 0.05 + \depthArrowPad})
      {$n$};
    \node[below, align=center, font=\footnotesize] at ({\fullR*0.5 + 2*\dx}, -0.55)
      {$\hat{\Tn{G}}_i$ (variable rank)};
  \end{scope}

  \node[above, font=\normalsize\bfseries] at ({(2.3 + \fullR)*0.5 + 2*\dx}, {4*\dy + \sliceH + 0.75})
    {t-SVDM-II};
  \node[above, font=\footnotesize] at ({(2.3 + \fullR)*0.5 + 2*\dx}, {4*\dy + \sliceH + 0.42})
    {rank $\rho_i \leq r$ varies per slice};

  \node[align=left, font=\footnotesize, gray!70!black]
    at ({2.75 + \fullR + 0.6}, {4*\dy + 0.85*\sliceH + 0.55}) {
      \begin{tikzpicture}
        \draw[wastebord, dashed, thin] (0,0) -- (0.35,0);
        \node[right, font=\footnotesize, gray!70!black] at (0.35,0) {discarded};
      \end{tikzpicture}
    };

\end{scope}

\node[font=\footnotesize, align=center, gray!60!black]
  at (5.5cm, {-\sliceH*0.0 - 0.2cm})
  {};

\end{tikzpicture}
    \caption{The output factor structure of t-SVDM-I (left) has uniform ranks across frontal slices which yield coarser error control, whereas t-SVDM-II (right) allows for variable frontal slice ranks $\rho_i$ which can adapt to approximation requirements.}
    \label{fig:tsvdm_output_updated}
\end{figure}
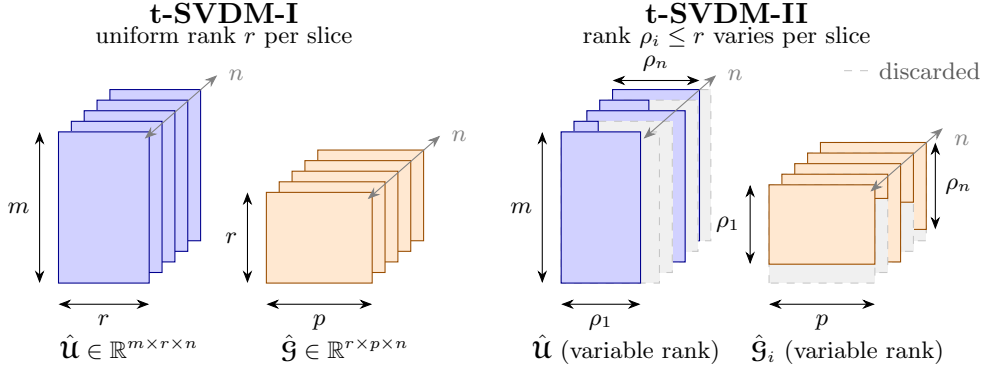

\subsection{Notation}\label{subsec:notation}
We use uppercase bold letters (e.g., $\Mx{M}$) to denote matrices and calligraphic letters (e.g., $\Tn{A}$) to denote tensors.
A three-way tensor $\Tn{A} \in \mathbb{R}^{m \times p \times n}$ has $n$ frontal slices $\Tn{A}_{:,:,i} \in \mathbb{R}^{m \times p}$ for $i = 1, \dots, n$.
The mode-$k$ unfolding $\Mx{A}_{(k)}$ arranges the mode-$k$ fibers of $\Tn{A}$ as columns of a matrix.
The mode-$k$ tensor times matrix product (TTM), written $\Tn{A} \ttm[k] \Mx{M}$ for $\Mx{M} \in \mathbb{R}^{J \times n_k}$, is defined by
\[
  \bigl(\Tn{A} \ttm[k] \Mx{M}\bigr)_{(k)} \;=\; \Mx{M} \, \Mx{A}_{(k)}.
\]
The result has the same shape as $\Tn{A}$ except that mode $k$ has size $J$ instead of $n_k$.

\paragraph{\starM{}-Product}
The \starM{}-product, introduced by Kilmer et al.~\cite{kilmer2021tensor}, is a tensor-tensor multiplication parameterized by an invertible matrix $\Mx{M}\in\mathbb{R}^{n\times n}$.
Given $\Tn{A}\in\mathbb{R}^{m\times p\times n}$ and $\Tn{B}\in\mathbb{R}^{p\times\ell\times n}$, the \starM{}-product $\Tn{C}=\Tn{A}\starM\Tn{B}\in\mathbb{R}^{m\times\ell\times n}$ is defined in three steps.
First, both operands are transformed to the $\Mx{M}$-domain by applying $\Mx{M}$ along mode~3,
\[
  \hat{\Tn{A}} = \Tn{A}\ttm[3]\Mx{M}, \qquad \hat{\Tn{B}} = \Tn{B}\ttm[3]\Mx{M}.
\]
Second, the $n$ pairs of frontal slices are multiplied independently in the transform domain,
\[
  \hat{\Tn{C}}_{:,:,i} \;=\; \hat{\Tn{A}}_{:,:,i}\,\hat{\Tn{B}}_{:,:,i}, \qquad i=1,\dots,n,
\]
where each product is an ordinary matrix multiplication.
Third, the result is mapped back to the original domain by applying $\Mx{M}^{-1}$ along mode~3,
\[
  \Tn{C} \;=\; \hat{\Tn{C}}\ttm[3]\Mx{M}^{-1}.
\]
When $\Mx{M}$ is a nonzero scalar multiple of an orthogonal matrix, $\Mx{M}^{-1}$ reduces to a (scaled) transpose, and the \starM{}-product inherits matrix-mimetic properties such as an associative and distributive algebra, a notion of orthogonality and identity tensors, and the Eckart-Young-style optimality results which are proved by Kilmer et al.~\cite{kilmer2021tensor}\footnote{Extensions to $4^\text{th}$ and higher order tensors are straightforward. Transformation to/from the $\Mx{M}$-domain occurs via TTMs with $\Mx{M}_i$/$\Mx{M}^{-1}_i$ for $i \ge 3$. A frontal slice is defined by fixing all indices except the first two for all the slice-wise operations. }.

\subsection{Tensor Singular Value Decomposition}\label{subsec:tsvd}
In this section, we provide a brief overview of the mathematical background and algorithms that underpin our work.
We focus on the tensor singular value decomposition (t-SVD) for compressing scientific datasets.
The t-SVD can be defined in many ways, but we focus on the \starM{} t-SVD framework developed by Kilmer et al.~\cite{kilmer2021tensor}, which is defined with respect to an invertible matrix $\Mx{M}$. 

Given an invertible matrix $\Mx{M}\in\mathbb{R}^{n\times n}$ and a 3-way tensor $\Tn{A}\in\mathbb{R}^{m\times p\times n}$, the \starM{} t-SVD of $\Tn{A}$ is the factorization $\Tn{A}=\Tn{U}\starM\Tn{S}\starM\Tn{V}$, where $\Tn{U}$ and $\Tn{V}$ are orthogonal tensors and $\Tn{S}$ is an f-diagonal tensor (i.e., diagonal frontal slices).
The key computational step is transforming $\Tn{A}$ to the $\Mx{M}$-domain by applying $\Mx{M}$ to mode~3,
\[
\hat{\Tn{A}} = \Tn{A} \ttm[3] \Mx{M},
\]
after which the frontal slices $\hat{\Tn{A}}_{:,:,i}$ are mutually decoupled and each can be independently factored by a standard thin matrix SVD.
The output factor tensors $\hat{\Tn{U}}$, $\hat{\Tn{S}}$, $\hat{\Tn{V}}$ are then mapped back to the original domain by applying $\Mx{M}^{-1}$ to mode~3.
The full \starM{} t-SVD is given in \Cref{alg:full-tsvdm}, and the adaptive truncated variant with relative error tolerance $\varepsilon$ is given in \Cref{alg:tsvdm2} (see \Cref{fig:tsvdm_output_updated}).

\paragraph{Choice of $\Mx{M}$.}
The transform matrix $\Mx{M}$ is an important decision point in the \starM{} framework.
Since $\Mx{M}$ determines the basis used to decouple the frontal slices, it directly affects the singular-value distribution and the resulting compression quality.
Prior work has considered fixed analytic transforms (e.g. DCT and DFT), data-dependent transforms computed from $\Tn{A}$ (e.g. singular vectors of the mode-3 unfolding), and learned transforms obtained by optimization.

Kilmer et al.~\cite{kilmer2021tensor} survey several fixed, analytic choices that inherit the Eckart-Young optimality guarantee of the \starM{} t-SVD framework.
The (normalized) discrete Fourier transform (DFT) recovers the original t-product of Kilmer and Martin and allows an $O(n\log n)$ application via the FFT, but yields complex-valued frontal slices in the transform domain.
The discrete cosine transform (DCT) is a real-valued alternative that can also be applied with an $O(n\log n)$ algorithm.

$\Mx{M}$ can also be tailored to the input tensor.
A simple data-driven choice discussed in~\cite{kilmer2021tensor} is to take $\Mx{M}$ as the (transposed) matrix of left singular vectors of the mode-3 unfolding $\Mx{A}_{(3)}$, which empirically yields better approximation at a given truncation rank than other fixed transforms.
However, this data-driven approach comes at the cost of an upfront SVD on the unfolding.

More recently, Newman and Keegan~\cite{newman2025optimal} have proposed a variable projection (VarPro) approach that jointly optimizes $\Mx{M}$ and the truncated factors.
The inner slicewise SVD problem is solved in closed form as a function of $\Mx{M}$ and projected out, leaving a reduced nonlinear optimization problem.
This produces transforms tailored to specific datasets and downstream tasks, at the cost of an iterative training step.
Our work treats $\Mx{M}$ as an arbitrary user-supplied invertible matrix, so any of the above choices can be utilized while leveraging the speedups achievable through shared-memory parallelization.

\begin{algorithm}
\caption{Full t-SVDM of $\Tn{A}$}
\label{alg:full-tsvdm}
\begin{algorithmic}[1]
\Require $\Tn{A}\in\mathbb{R}^{m\times p\times n}$, invertible $\Mx{M}\in\mathbb{R}^{n\times n}$
\Ensure $\Tn{U}$, $\Tn{S}$, $\Tn{V}$ such that $\Tn{A}=\Tn{U}\starM\Tn{S}\starM\Tn{V}^{\intercal}$
\State $\hat{\Tn{A}} \gets \Tn{A} \ttm[3] \Mx{M}$
\For{$i = 1,\dots,n$}
  \State $[\hat{\Tn{U}}_{:,:,i},\; \hat{\Tn{S}}_{:,:,i},\; \hat{\Tn{V}}_{:,:,i}] \gets \operatorname{svd}(\hat{\Tn{A}}_{:,:,i})$
\EndFor
\end{algorithmic}
\end{algorithm}

\begin{algorithm}
\caption{t-SVDM-II with relative error tolerance $\varepsilon$}
\label{alg:tsvdm2}
\begin{algorithmic}[1]
\Require Tensor $\Tn{A}\in\mathbb{R}^{m\times p\times n}$, $\Mx{M}\in\mathbb{R}^{n\times n}$ a nonzero multiple of an orthogonal matrix, relative error tolerance $\varepsilon\in(0,1)$
\Ensure Per-face truncation levels $\rho=(\rho_1,\dots,\rho_n)$ and truncated factors $\{\hat{\Tn{U}}_{:,1:\rho_i,i}$, $\hat{\Mx{G}}_{\rho_i}\}_{i=1}^n$
\State Compute the t-SVDM of $\Tn{A}$
\State $v \gets \text{concatenate}\!\left( (\hat{\Tn{S}}_{j,j,i})^2 \text{ for all } i,j \right)$
\State $v \gets \text{sort}(v,\text{ ascending})$
\State Let $w$ be the cumulative-sum vector, i.e., $w_k=\sum_{\ell=1}^{k} v_\ell$
\State Find the largest index $J$ such that $w_J / \|\hat{\Tn{S}}\|_F^2 < \varepsilon^2$
\State $\tau \gets v_J$ \Comment{largest singular value to be discarded}
\For{$i = 1,\dots,n$}
  \State $\rho_i \gets$ number of singular values of $\hat{\Tn{A}}_{:,:,i}$ greater than $\tau$
  \State Keep only $\hat{\Tn{U}}_{:,1:\rho_i,i}$ and form $\hat{\Mx{G}}_{\rho_i} \gets \hat{\Tn{S}}_{1:\rho_i,1:\rho_i,i}\,\hat{\Tn{V}}^{\intercal}_{:,1:\rho_i,i}$
\EndFor
\end{algorithmic}
\end{algorithm}

\subsection{Optimality of \starM{} t-SVD}\label{subsec:optimality}
The primary advantage of the \starM{} t-SVD is that it admits an Eckart-Young-like optimality guarantee whereas  established tensor decompositions such as the CANDECOMP/PARAFAC (CP)~\cite{hitchcock1927expression,carroll1970analysis,harshman1970foundations}, Tucker~\cite{tucker1966some}, Tensor-Train SVD~\cite{oseledets2011tensor}, and the higher-order SVD (HOSVD)~\cite{delathauwer2000multilinear} families do not (see also the survey by Kolda and Bader~\cite{kolda2009tensor}).
For matrices, the Eckart-Young theorem states that truncating the SVD to its leading rank-$k$ singular values and vectors yields the best rank-$k$ approximation in the Frobenius (and 2-) norm~\cite{eckart1936approximation,mirsky1960symmetric}.
This single property is what makes the matrix SVD the primary tool for dimensionality reduction and lossy compression.
Generalizing this guarantee to multiway data has been a long-standing open problem.

For the CP decomposition, determining the tensor rank is NP-hard in general, and the factor matrices are not required to be orthogonal or even full rank~\cite{kilmer2021tensor}.
As a result, no provably optimal truncation policy is known.
Practical implementations often rely on iterative alternating-least-squares schemes that may converge slowly, are sensitive to initialization, and offer no global-optimality guarantees on the resulting approximation.
The Tucker and HOSVD decompositions do produce factor matrices with orthonormal columns, but the core tensor is typically dense and does not yield an optimal low-rank approximation.
Instead, the truncated HOSVD is only \emph{quasi-optimal}: for a $d$-way tensor $\Tn{A}$, an HOSVD-truncated approximation $\Tn{B}_{\Mx{k}}$ satisfies
\[
  \|\Tn{A} - \Tn{B}_{\Mx{k}}\|_F \;\le\; \sqrt{d}\,\|\Tn{A} - \Tn{B}^{\star}\|_F,
\]

where $\Tn{B}^{\star}$ is the optimal multilinear-rank-$\Vc{k}$ approximation~\cite{delathauwer2000bestrank}.
The $\sqrt{d}$ factor compounds with tensor order, so the gap grows as the data become more high-dimensional.

In contrast, when $\Mx{M}\in\mathbb{R}^{n\times n}$ is
 a nonzero scalar multiple of an orthogonal matrix, Kilmer et al.~\cite{kilmer2021tensor} prove that both the t-SVDM-I (uniform-rank) and t-SVDM-II (error-tolerance, variable-rank) truncations are \emph{exactly} Eckart-Young optimal in the Frobenius norm under the framework induced by $\starM$.
This work showed that truncating the singular values of the transformed slices directly produces the best lower-rank representation.
Combined with the fact that the slice-wise SVDs in the $\Mx{M}$-domain are decoupled (and easily parallelizable), the \starM{} framework provides an attractive option for compressing large scientific datasets where low-rank optimality and computational efficiency are important.

\subsection{Shared-Memory Parallel \starM{} t-SVD}\label{subsec:motivation}
Despite the Eckart-Young-like optimality results, the practical impact of the \starM{} t-SVD is limited by the available software.
The reference implementations accompanying~\cite{kilmer2021tensor} and the broader \starM{} work are written in productivity-oriented languages, such as  MATLAB and pure Python.
While these implementations are excellent for small scale experimentation with the \starM{} framework, they are not directly suitable for large-scale data compression tasks on modern multicore architectures.
To our knowledge, no shared-memory parallel, high-performance implementation of the \starM{} t-SVD currently exists.
This is in contrast to the software maturity of the matrix SVD (e.g. LAPACK and ScaLAPACK), CP, and Tucker decompositions, for which both shared-memory and distributed-memory parallel libraries are available~\cite{phipps2019software,smith2016medium,austin2016parallel,ballard2020tuckermpi}.

This work closes the parallel algorithms design and software gap by developing the \texttt{pystarm} software package, a shared-memory parallel, high-performance implementation of the \starM{} t-SVD that exposes parallelism in the tensor-times-matrix and the slicewise SVD operations.
We utilize threaded BLAS and LAPACK kernels through Intel MKL, and provide a Python interface so that performance and productivity are simultaneously achievable for domain scientists.

\section{Design and Implementation}
\label{sec:implementation}
We implement the two algorithms described in \Cref{sec:background} in C++ using Intel's Math Kernel Library (MKL) for dense linear algebra, with Python bindings provided via pybind11.
Throughout, we restrict $\Mx{M}$ to be orthonormal, so $\Mx{M}^{-1} = \Mx{M}'$, which eliminates explicit matrix inversion.
Recent work has extended the t-SVDM framework to other choices of $\Mx{M}$, including non-orthonormal transformations~\cite{kernfeld2015tensor,keegan2025projected}.
We focus on orthonormal choices of $\Mx{M}$ which provide optimality guarantees while reducing computational complexity when backward transformations of \starM{} t-SVD outputs are required.
All tensor and matrix data are stored in column-major order.
This choice is consequential: in column-major layout the $m \times p$ frontal slices of a 3-way tensor $\Tn{A} \in \mathbb{R}^{m \times p \times n}$ occupy contiguous, non-overlapping memory regions, so each slice can be addressed as a matrix without data movement or transposition.
Both algorithms decompose into two main computational kernels---tensor-times-matrix multiplication and slice-wise SVD---which we describe in turn before addressing the specifics of each variant.
All microbenchmarks reported in this section for the two computational kernels were performed on a dual socket Intel Xeon CPU machine with 52 cores per socket.

\subsection{Tensor-Times-Matrix}
\label{subsec:ttm}

The TTM $\hat{\Tn{A}} = \Tn{A} \ttm[k] \Mx{M}$ is the workhorse of both algorithms.
For a 3-way tensor it applies $\Mx{M}$ to mode~3 to enter the transform domain; for a $d$-way tensor it applies $\Mx{M}$ to every mode $k > 2$ (with potentially a different transform per mode).
The first two modes are never contracted by the algorithm, so we only need TTM kernels for modes $3,\dots,d$.
Among these, mode $d$ is the \emph{last mode} of the tensor and modes $3,\dots,d-1$ are \emph{middle modes}; the structure of the computation differs in the two cases.

For a TTM on the last mode of a $d$-way tensor, the column-major layout exposes the tensor as a single $\bigl(\prod_{i<d} n_i\bigr) \times n_d$ matrix; right-multiplying it by $\Mx{M}' \in \mathbb{R}^{n_d \times J}$ produces the correctly contracted result and is a single level-3 BLAS GEMM call~\cite{Ballard_Kolda_2025}.
For a 3-way tensor this is the only TTM that arises ($k = d = 3$), so MKL's multi-threaded GEMM handles all parallelism automatically.

For a TTM on a middle mode $k$ with $2 < k < d$, the mode-$k$ fibers are not contiguous, so the operation must be expressed as a collection of GEMM calls.
Let $M_k = \prod_{i < k} n_i$ and $P_k = \prod_{i > k} n_i$.
The tensor can be viewed as $P_k$ contiguous submatrices of shape $M_k \times n_k$, each to be right-multiplied by $\Mx{M}' \in \mathbb{R}^{n_k \times J}$ to yield an $M_k \times J$ result~\cite{Ballard_Kolda_2025}.
Middle-mode TTMs only arise once the tensor has more than three modes, but then they dominate: in the 6-way \emph{ncep-air-6} dataset (\Cref{subsec:climate}), three of the four required TTMs (modes 3, 4, and 5) are middle-mode contractions.
The efficient parallelization of these middle-mode TTMs is therefore an important design choice for a high-performance shared-memory implementation of the \starM{} t-SVD.

\subsubsection{Multithreaded TTM Design Choices}
We identify three primary choices to implement multithreaded TTM for the middle modes.

\paragraph*{Choice 1: Batched GEMM}
We can use MKL's batched strided GEMM interface, in which a single MKL call performs all $P_k$ multiplications such that: the input stride between successive submatrices is $M_k n_k$, the output stride is $M_k J$, and the stride for $\Mx{M}$ is zero so the same matrix is reused across all batches.
This allows MKL's scheduling layer to optimize across the batch, reducing per-call overhead compared to an equivalent loop over individual DGEMM calls.
For the mode-3 TTM that appears in both algorithms on a 3-way tensor, $P_k = 1$ and $M_k = m p$, so the batched call degenerates to a single DGEMM that right-multiplies the $(m p) \times n$ matrix view of the tensor by $\Mx{M}' \in \mathbb{R}^{n \times J}$, yielding an $(m p) \times J$ result.
We call this the \emph{batched} variant of TTM.

\paragraph*{Choice 2: Loop over multithreaded GEMM}
We can call the individual GEMM operations one after another sequentially and use optimized multithreaded GEMM kernels for each of the individual GEMMs.
We call this the \emph{loop} variant of TTM.

\paragraph*{Choice 3: Parallel for with sequential GEMM}
We can lift the parallelism from MKL multithreading to OpenMP multithreading by leveraging a parallel for over the batch of GEMM and use sequential MKL GEMM calls.
These calls can be issued simultaneously without synchronization.
We call this the \emph{parfor} variant of TTM.

\subsubsection{Performance of three TTM designs}
\begin{figure}[H]
  \centering
  \includegraphics[width=0.65\textwidth,keepaspectratio]{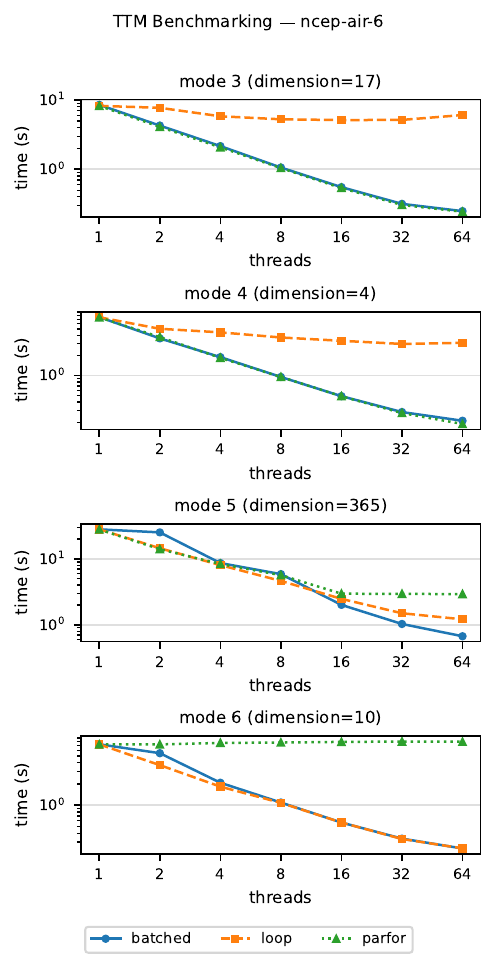}
  \caption{TTM performance of three variants (\emph{batched} vs.\ \emph{loop} vs.\ \emph{parfor}) for the \emph{ncep-air-6} dataset across thread counts and TTM modes.}
  \label{fig:benchmark-ttm-ncep-air-6}
\end{figure}

We compare the performance of the three design choices of TTM on the \emph{ncep-air-6} tensor discussed in \Cref{subsec:climate}.
It is a 6-way tensor and exercises both the middle-mode (modes 3, 4, 5) and last-mode (mode 6) cases that the algorithm actually invokes.
We show the comparison in \Cref{fig:benchmark-ttm-ncep-air-6}.

We observe that the \emph{batched} variant works best independent of which mode the TTM operation is performed on.
The \emph{parfor} variant suffers for the last mode.
For TTM on the last mode, the \emph{parfor} variant reduces to a single sequential GEMM call, hence it does not scale at all with increasing thread count.
The \emph{loop} variant suffers for the first few modes.
As explained above, TTM on a middle mode reduces to multiple GEMM calls.
For the first few modes, the \emph{loop} variant reduces to multiple calls of multithreaded GEMM each having a small workload.
Hence, the scalability suffers at higher thread counts.
For the remainder of this paper we report experiments with only the \emph{batched} variant of TTM as it is the clear winner among all three design choices.

\subsection{Slice-wise SVD}
\label{subsec:slicewise-svd}
After transforming to the $\Mx{M}$-domain, both algorithms require computing the matrix SVD of each frontal slice $\hat{\Tn{A}}_{:,:,i}$ independently.
Because LAPACK's SVD routine overwrites its input, each slice is materialized as a fresh copy before factorization.

We support two factorization modes.
The first computes the full thin SVD of each slice, yielding factor tensors $\hat{\Tn{U}} \in \mathbb{R}^{m \times r \times n}$ and $\hat{\Tn{V}} \in \mathbb{R}^{p \times r \times n}$ (where $r = \min(m, p)$) together with a matrix of singular values.
The second computes only the leading $k$ singular triplets via a truncated SVD, avoiding the cost of forming the full factorization when $k \ll r$.

\subsubsection{Multithreaded Design Choices}
The $n$ slice SVDs are independent, presenting two natural parallelization strategies.

\paragraph*{Choice 1: Sequential slices, parallel SVD}
Each LAPACK call is itself multi-threaded, exploiting intra-slice parallelism through MKL's internal threading.

\paragraph*{Choice 2: Parallel slices, sequential SVD}
In the second, a parallel loop distributes slices across threads, with each thread calling a single-threaded LAPACK routine.

The first strategy is preferable when the slices are large relative to the number of slices ($m, p \gg n$), so each SVD has enough work to keep many threads busy; the second is preferable when there are many small slices ($n \gg m, p$), where inter-slice parallelism offers more opportunity than intra-slice parallelism.

\begin{figure}[H]
	\centering
	\includegraphics[width=0.65\textwidth,keepaspectratio]{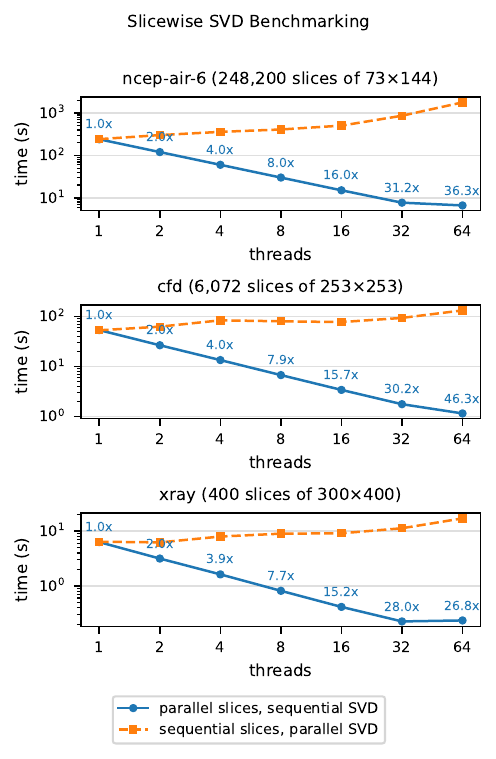}
	\caption{Slice-wise SVD wall time (parallel slices, sequential SVD vs.\ sequential slices, parallel SVD) for the NCEP-Air 6-way, CFD, and X-ray Crystallography datasets.}
	\label{fig:benchmark-svd}
\end{figure}

\subsubsection{Performance of the designs}
In \Cref{fig:benchmark-svd} we show a microbenchmark demonstrating the performance of the two design choices of slicewise SVD on the three practical tensors described in \Cref{sec:application}.
We observe that the \emph{parallel-slices-with-sequential-svd} variant gives almost linear scalability over increasing thread count.
In contrast, the \emph{sequential-slices-with-parallel-svd} variant does not scale.
We believe that this is due to the small sizes of the SVD where parallel overhead dominates.
Given the complexity of parallelizing the SVD computation, we expect the  \emph{parallel-slices-with-sequential-svd} to yield better performance for most applications.


\subsection{Full t-SVDM and \tsvdmi}
\paragraph*{Full t-SVDM}
\Cref{alg:full-tsvdm} shows the full t-SVDM algorithm.
For simplicity, the algorithm is shown for a three-way tensor.
This variant computes all singular values and vectors of each frontal slice.
In contrast,  \tsvdmi computes a truncated SVD with fixed rank-$r$ for each frontal slice.
This variant enables predictable performance in each SVD call and reduces the overhead of computing variable rank SVDs and the complexity of determining ranks adaptively.
The full SVD uses the \texttt{dgesvd} LAPACK call, whereas the truncated SVD calls \texttt{dgesvdx}.

\subsection{\tsvdmii}
\label{subsec:tsvdm2}
\begin{figure}[H]
	\centering
	\includegraphics[width=0.65\textwidth,keepaspectratio]{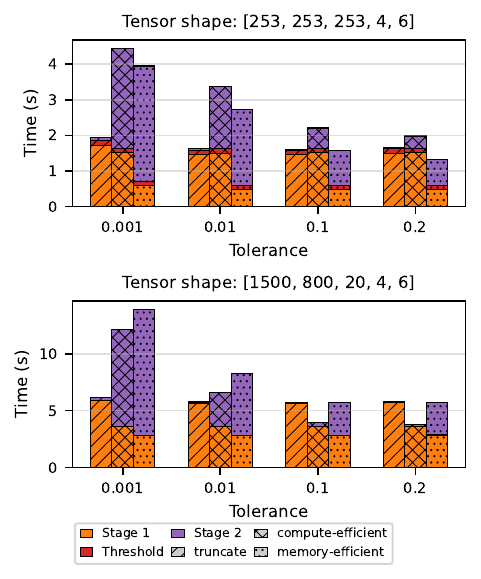}
	\caption{Breakdown times for the different t-SVDM-II strategies. Depending on the time and memory constraints, any of the three methods can be preferred.}
	\label{fig:benchmark-svd-bkd}
\end{figure}

The adaptive truncated variant (\Cref{alg:tsvdm2}) introduces per-slice truncation ranks $\rho_i$ that vary across slices and are not known until after the singular values have been inspected globally. This data dependency rules out the simple composition used for t-SVDM-I and motivates three strategies with different trade-offs between computational redundancy and memory efficiency.

\subsubsection{Multithreaded Design Choices}
All strategies consist of three stages: (1) compute the singular values of all slices, (2) determine the threshold $\tau$, and (3) compute the final factorization.

\paragraph{Choice 1: Truncate}
The truncation strategy is a straightforward way to compute the t-SVDM-II factorization. We compute the full t-SVDM (\Cref{alg:full-tsvdm}) factorization, determine the threshold $\tau$ from all $n \cdot r$ singular values, and discard the singular vectors corresponding to values below $\tau$.
This is algorithmically simple and reuses the same slice-wise SVD kernel without modification, but pays the full cost of computing all singular vectors only to discard most of them when compression is aggressive.
The \emph{truncate} variant consumes the most intermediate memory of all three strategies storing up to $3\times$ the memory footprint of the input data.

\paragraph{Choice 2: Compute-efficient.}
In the second strategy, a first pass computes only the singular values of each slice, without forming any singular vectors.
The SVD algorithm consists of multiple steps (a possible QR factorization and bidiagonal reduction) before computing the singular values and vectors. We cache these intermediate entities.
The global threshold $\tau$ and per-slice ranks $\rho_i$ are determined from this pass.
A second pass then computes only the leading $\rho_i$ singular triplets for each slice, via a truncated SVD. We utilize the cached intermediaries to speed up the second SVD computation.
All output buffers are formed to the exact required ranks before the second pass.
The compressed factorization is stored in a variable-rank format in which the $i$-th slice's factor matrices occupy exactly $m \times \rho_i$ and $p \times \rho_i$ entries with no padding, as opposed to the dense $m \times r \times n$ and $p \times r \times n$ tensors produced by t-SVDM-I.
This representation accurately reflects the storage cost of the compressed result and avoids allocating memory for singular components that are not retained.
When target ranks are small, the savings in the second stage outweigh the cost of reading each slice twice.
This \emph{compute-efficient} version of t-SVDM-II reduces the memory footprint by half when compared to the \emph{truncate} strategy.
It also performs the fewest floating-point operations of all three strategies for most cases.


\paragraph{Choice 3: Memory-efficient}
Finally, we consider the most memory efficient strategy, where we do not save any intermediaries from the preceding strategy.
The second SVD computation could possibly take longer than the \emph{compute-efficient} strategy but we also save on the memory bandwidth needed in moving and copying the intermediaries.
This strategy does not use any more space than that needed for holding the input and output buffers in memory.
We call it the \emph{memory-efficient} variant.

\subsubsection{Performance of the three variants}
\Cref{fig:benchmark-svd-bkd} shows the runtime breakdown of the different strategies on two data tensors for various relative-error tolerances. All times are averaged over 5 runs. The first tensor is the CFD data ($253\times253\times253\times4\times6$, $\approx 3$~GiB) and the second is a larger tensor ($1500\times800\times20\times4\times6$, $\approx 4.3$~GiB) with a similar singular value distribution.
No clear winner exists among the three strategies.
The \emph{truncate} strategy is most insensitive to the tolerance parameter as it computes the full t-SVDM every time. Its runtime is fairly consistent over all tolerances with the bulk of the time in the Stage 1 SVD.
The \emph{compute-efficient} and \emph{memory-efficient} strategies show more differing behavior. The Stage 2 times for \emph{compute-efficient} are always lower than those of \emph{memory-efficient}, but the gain may not always be sufficient to overcome the time spent in moving the cached intermediaries in memory during the first stage.
The Stage 1 SVD calls the \texttt{dgesvd} kernel which employs the implicit zero-shift QR algorithm for its bidiagonal SVD (\texttt{dbdsqr}), while the truncated SVD (\texttt{dgesvdx}) calls the bisection plus inverse iteration algorithm (\texttt{dbdsvdx}). Our studies showed an $8\times$ performance difference between these methods, with \texttt{dbdsqr} outperforming \texttt{dbdsvdx}. This explains the relative competitiveness of the \emph{truncate} strategy.
Since we are concerned with data compression, we will use the \emph{memory-efficient} strategy for all t-SVDM-II benchmarks.


\section{Application}
\label{sec:application}


In this section we demonstrate the practical utility of our work on three scientific datasets that are prohibitively large for currently available \tsvdmi and \tsvdmii implementations: atmospheric reanalysis data (\Cref{subsec:climate}), computational fluid dynamics data (\Cref{subsec:cfd}), and X-ray diffuse scattering data (\Cref{subsec:xray}). While tensor decompositions have many applications, we focus on compression as it provides a direct and quantifiable measure of the quality of the decomposition. For each dataset we evaluate compression quality and strong scaling performance.

\subsection{Climate Data (NCEP-Air)}
\label{subsec:climate}
\subsubsection{Dataset}

We use 10 years (1948--1957) of air temperature data from the NCEP/NCAR Reanalysis~\cite{kalnay1996ncep} at all 17 standard pressure levels, representing altitude.
We represent the data as a 6-way tensor of shape $73 \times 144 \times 17 \times 4 \times 365 \times 10$ (\texttt{ncep-air-6}), where the first three modes are latitude, longitude, and pressure level, and the time axis is split into time-of-day ($4$), day-of-year ($365$), and year ($10$).
February~29 observations are dropped from leap years to produce a uniform 365-day calendar, giving approximately 19.46\,GB as \texttt{float64}.
Compared to a flat 4-way representation (\texttt{ncep-air-4}, shape $73 \times 144 \times 17 \times 14{,}612$), in our experiments we observed that this reshaping achieves nearly identical compression quality while reducing TTM FLOPs by approximately $40\times$, as each time-mode GEMM operates on a much smaller dimension.

%
%
%
%

\subsubsection{Compression Quality}

\begin{figure}[H]
  \centering
  \includegraphics[width=0.65\textwidth]{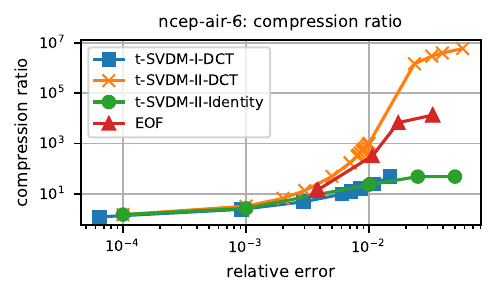}
  \caption{Compression ratio of the \textit{ncep-air-6} tensor for different algorithms.}
  \label{fig:ncep-air-compression}
\end{figure}

\Cref{fig:ncep-air-compression} compares compression ratio across methods on the \textit{ncep-air-6} tensor. \tsvdmii-DCT achieves higher compression than \tsvdmi-DCT at every error level, as per-slice variable truncation allows \tsvdmii to discard many slices entirely. Both tensor methods outperform EOF~\cite{hannachi2007eof}, the standard matrix-based compression in climate science, consistent with the theoretical results of~\cite{kilmer2021tensor}. The DCT transform further outperforms the identity transform, as it concentrates energy into fewer singular values per slice. For the remaining experiments we focus on comparing \tsvdmi and \tsvdmii with the DCT transformation.

\subsubsection{Strong Scaling}
\begin{figure}[H]
  \centering
  \includegraphics[width=0.65\textwidth]{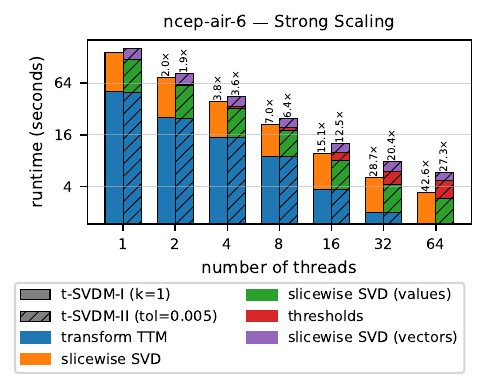}
  \caption{Strong scaling of both \tsvdmi and \tsvdmii for the \textit{ncep-air-6} data with DCT transformation. \tsvdmii is run with tolerance 0.005 (compression ratio $\approx$48.9), and \tsvdmi is run with rank $k=1$ (compression ratio $\approx$48.2).}
  \label{fig:ncep-scaling}
\end{figure}

\Cref{fig:ncep-scaling} shows strong scaling of \tsvdmi ($k=1$, compression ratio $\approx48.2\times$) and \tsvdmii (tol=$0.005$, compression ratio $\approx48.9\times$) on the \textit{ncep-air-6} tensor. \tsvdmi achieves $42\times$ speedup from 1 to 64 threads (147s $\to$ 3.5s) and \tsvdmii achieves $27\times$ (160s $\to$ 5.9s). For \tsvdmii, singular value computation dominates at low thread counts, while singular vector computation is relatively cheap --- due to high compressibility, most slices are dropped entirely after thresholding.


%
%

\subsubsection{Extreme Event Preservation}

\begin{figure}[H]
  \centering
  \includegraphics[width=0.65\textwidth]{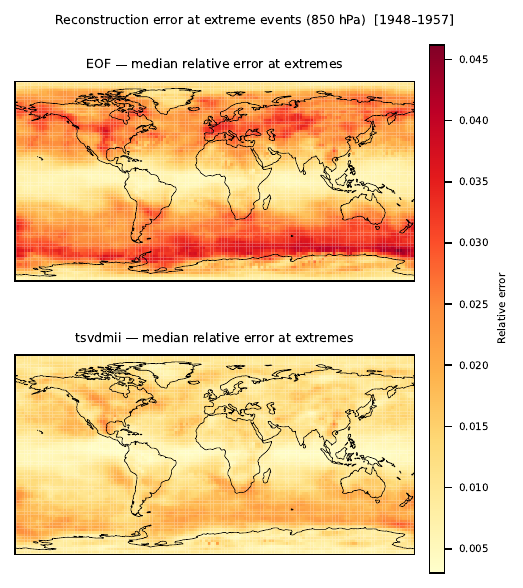}
  \caption{Median pointwise relative error at extreme temperature events (850\,hPa)
           for EOF (top) and \tsvdmii-DCT (bottom), at a matched compression ratio of $\sim$320$\times$.
           Lower is better.}
  \label{fig:ncep-air-extremes}
\end{figure}

\paragraph{EOF vs.\ \tsvdmii at extremes.}
A known limitation of EOF-based compression in climate science is that it can smooth extreme weather events~\cite{jiang2020pca}.
\Cref{fig:ncep-air-extremes} compares the median pointwise relative error at extreme temperature events (top and bottom 0.5\% of the time series at each grid point, at 850\,hPa — a pressure level that approximates surface air temperature) for EOF and \tsvdmii-DCT at a matched compression ratio of approximately 320$\times$.
For each spatial grid point, the error is computed over all time steps in the dataset and aggregated as the median across extreme events.
\tsvdmii-DCT achieves lower reconstruction error at extremes across most of the globe, demonstrating that tensor-based compression better preserves physically significant rare events.

\subsection{CFD}
\label{subsec:cfd}

\subsubsection{Dataset}

\begin{figure}[H]
  \centering
  \includegraphics[width=0.65\textwidth]{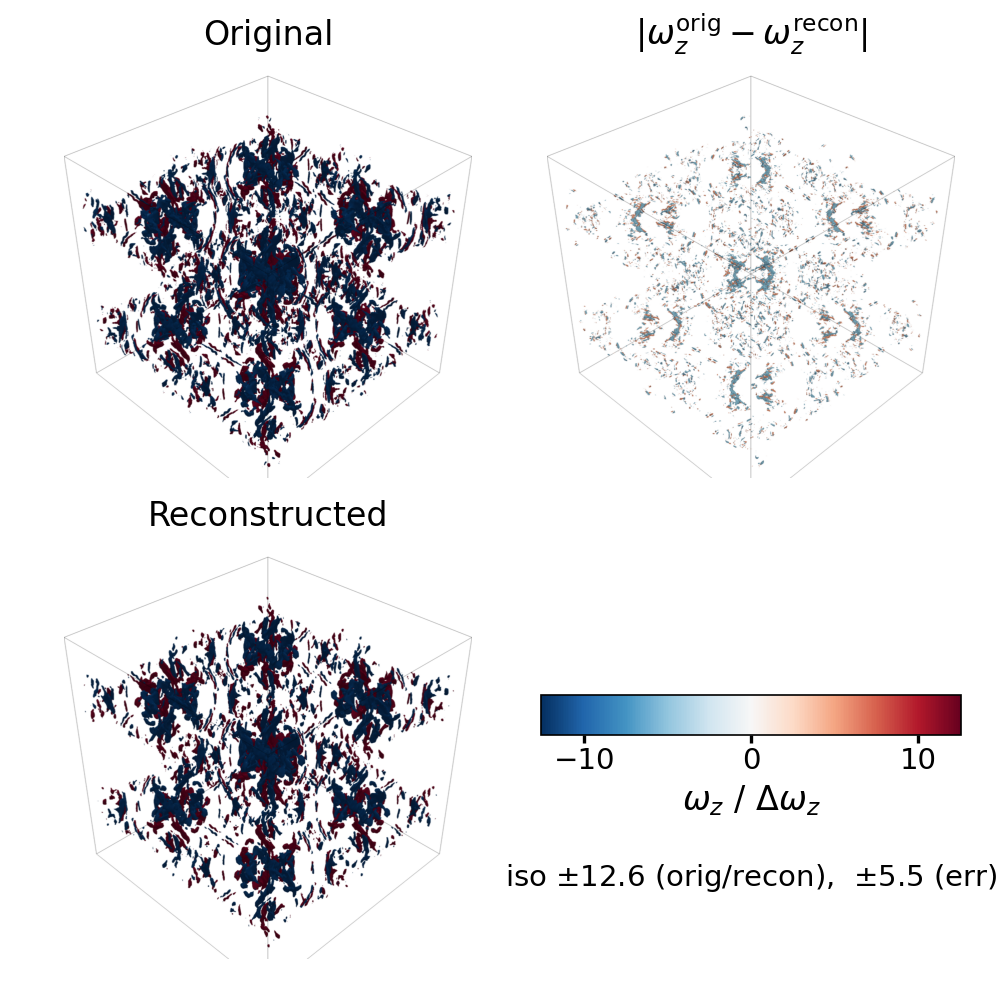}
  \caption{Iso-surfaces of the $z$-component of vorticity, $\omega_z = \partial v/\partial x - \partial u/\partial y$, for the Taylor--Green Vortex flow: original (top-left), compressed reconstruction (bottom-left), and pointwise error (top-right). Reconstruction is via \tsvdmii-DCT at tolerance $10^{-1}$ (compression ratio $\approx16\times$). Red and blue denote positive and negative values.}
  \label{fig:cfd-vorticity}
\end{figure}

We use a computational fluid dynamics (CFD) dataset generated from a direct numerical simulation of the Taylor--Green Vortex (TGV)~\cite{barwey2025mesh}, a standard benchmark for turbulent flow, solved using NekRS~\cite{fischer2022nekrs}. The simulation data is assembled into a 5-way tensor of shape $253 \times 253 \times 253 \times 4 \times 6$ (approximately 3\,GiB as \texttt{float64}), where the first three modes index the structured spatial grid, the fourth mode indexes four solution fields (velocity components $u, v, w$ and pressure $p$), and the last mode indexes six flow snapshots.
A visualization of one snapshot is shown in \Cref{fig:cfd-vorticity}.


\subsubsection{Compression Quality}
\begin{figure}[H]
  \centering
  \includegraphics[width=0.65\textwidth]{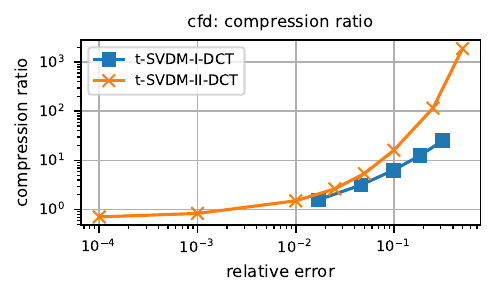}
  \caption{Compression ratio of the \textit{cfd} tensor for different algorithms.}
  \label{fig:cfd-compression}
\end{figure}

\Cref{fig:cfd-compression} shows the compression ratio of \tsvdmi and \tsvdmii on the \textit{cfd} tensor. At tight tolerances (relative error $\leq 0.05$), the two algorithms achieve comparable compression ratios --- for example, at relative error $\approx 0.05$, \tsvdmi ($k=20$) achieves $6.3\times$ while \tsvdmii achieves $5.4\times$. However, as the tolerance is relaxed, \tsvdmii pulls significantly ahead: at relative error $\approx 0.1$, \tsvdmii achieves $16\times$ compression compared to $6.3\times$ for \tsvdmi, and the gap grows further at higher tolerances. \Cref{fig:cfd-vorticity} shows the reconstructed vorticity field at tolerance $10^{-1}$, confirming that the compressed reconstruction closely matches the original.



\subsubsection{Strong Scaling}

\begin{figure}[H]
  \centering
  \includegraphics[width=0.65\textwidth]{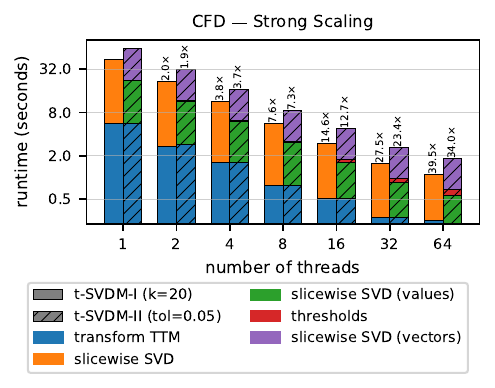}
  \caption{Strong scaling of both \tsvdmi and \tsvdmii for the \textit{cfd} data with DCT transformation. \tsvdmii is run with tolerance 0.05 (compression ratio $\approx$5.4), and \tsvdmi is run with rank $k=20$ (compression ratio $\approx$6.3).}
  \label{fig:cfd-scaling}
\end{figure}

\Cref{fig:cfd-scaling} shows strong scaling of \tsvdmi ($k=20$, compression ratio $\approx6.3\times$) and \tsvdmii (tol=$0.05$, compression ratio $\approx5.4\times$) on the \textit{cfd} tensor. \tsvdmi achieves a $39\times$ speedup from 1 to 64 threads (43s $\to$ 1.1s), and \tsvdmii achieves $34\times$ (61s $\to$ 1.8s). \tsvdmii splits the SVD into two passes (singular values first, then singular vectors for retained singular values), which results in slightly weaker scaling. Unlike the \textit{ncep-air-6} tensor, singular vector computation accounts for a significant fraction of \tsvdmii runtime --- at 1 thread it takes $\approx39$s, comparable to the full SVD cost of \tsvdmi.


\subsection{X-ray Crystallography}
\label{subsec:xray}

\subsubsection{Dataset}

\begin{figure}[H]
  \centering
  \includegraphics[width=0.65\textwidth]{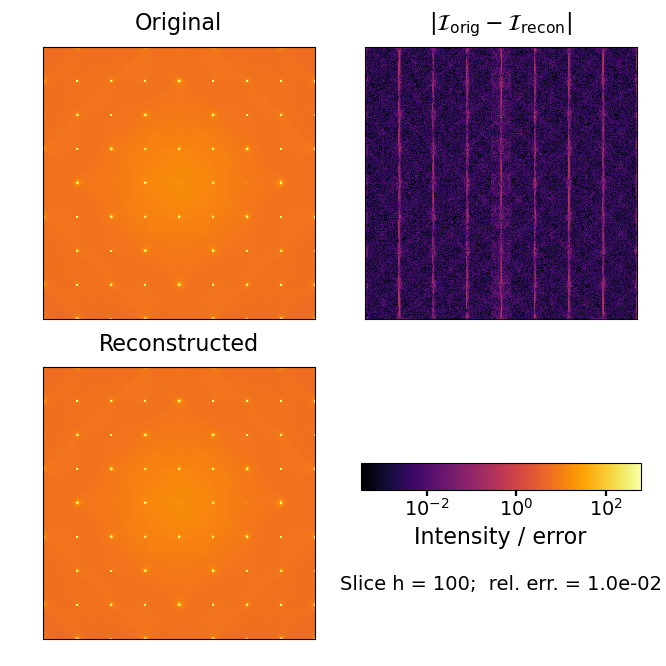}
  \caption{Two-dimensional slice through the X-ray diffuse-scattering volume: original (top-left), compressed reconstruction (bottom-left), and pointwise absolute error (top-right). Reconstruction is via \tsvdmii-DCT at tolerance $10^{-2}$ (compression ratio $\approx1.51\times$). A shared logarithmic colormap is used across all panels.}
  \label{fig:xray-slices}
\end{figure}

We use a 3D X-ray diffuse scattering dataset measured on a MoVO\textsubscript{2} single crystal~\cite{kuang2025sparse}, stored as a 3-way tensor of shape $300 \times 400 \times 400$ in \texttt{float64} (approximately 366\,MB), where each mode corresponds to one of the three orthogonal axes of the volumetric reciprocal-space measurement. A representative slice through the volume is shown in \Cref{fig:xray-slices}.


\subsubsection{Compression Quality}

\begin{figure}[H]
  \centering
  \includegraphics[width=0.65\textwidth]{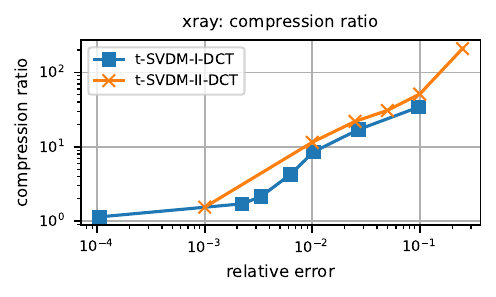}
  \caption{Compression ratio of the \emph{xray} tensor for different algorithms.}
  \label{fig:xray-compression}
\end{figure}

\Cref{fig:xray-compression} shows the compression ratio of \tsvdmi and \tsvdmii on the \textit{xray} tensor. \tsvdmii achieves higher compression than \tsvdmi at all error levels. \Cref{fig:xray-slices} shows the reconstructed slice at tolerance $10^{-2}$, confirming that the reconstruction closely matches the original.


\subsubsection{Strong Scaling}

\begin{figure}[H]
  \centering
  \includegraphics[width=0.65\textwidth]{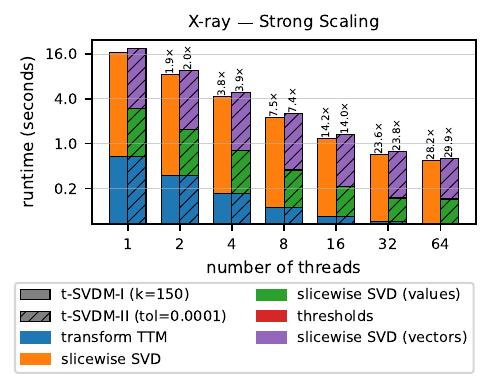}
  \caption{Strong scaling of both \tsvdmi and \tsvdmii for the \emph{xray} data with DCT transformation. \tsvdmii is run with tolerance 0.0001 (compression ratio $\approx$1.19), and \tsvdmi is run with rank $k=150$ (compression ratio $\approx$1.14).}
  \label{fig:xray-scaling}
\end{figure}

\Cref{fig:xray-scaling} shows strong scaling of \tsvdmi ($k=150$, compression ratio $\approx1.14\times$) and \tsvdmii (tol=$10^{-4}$, compression ratio $\approx1.19\times$) on the \textit{xray} tensor. Both algorithms scale similarly --- \tsvdmi achieves $28\times$ speedup from 1 to 64 threads (16.7s $\to$ 0.59s) and \tsvdmii achieves $30\times$ (18.7s $\to$ 0.63s). In both cases, slicewise SVD dominates the runtime.


\section{Conclusions and Future Work}\label{sec:conc}

We present the first high-performance implementations of tensor decompositions in the $\starM{}$-framework.
Our shared-memory parallel implementation takes advantage of the recent batch/strided interfaces developed for these linear algebra kernels.
We achieve good scaling up to 64 threads on three real-world datasets from different scientific applications.
Our case study on the climate dataset also highlights the utility of maintaining the multidimensionality of the dataset by preserving physically relevant rare events.


Our initial investigations open up many future directions for research.
Supporting GPUs in the library is an immediate next step. The embarrassingly parallel nature of the slice-wise operations would greatly benefit from this development. The irregular data access pattern due to varying ranks in t-SVDM-II also poses some difficulty.
Developing distributed-memory parallel algorithms is a more involved next step. Overcoming the difference in data layout preferences for TTM and SVD stage without an explicit transpose operation is an immediate concern.
Finally, we utilize explicit matrix multiplications in our TTM stage. Supporting operator transforms, for transforming the modes, is another avenue for future work. 
 

\FloatBarrier
\bibliographystyle{plain}
\bibliography{references}

@manual{Lu18,
  author       = {Lu, Canyi},
  title        = {Tensor-Tensor Product Toolbox},
  organization = {Carnegie Mellon University},
  month        = {June},
  year         = {2018},
  note         = {\url{https://github.com/canyilu/tproduct}},
  }

@article{kuang2025sparse,
  title   = {Recovering Sparse {DFT} from Missing Signals via Interior Point Method on {GPU}},
  author  = {Kuang, Wei and Rao, Vishwas and Montoison, Alexis and Pacaud, Fran\c{c}ois and Anitescu, Mihai},
  journal = {arXiv preprint arXiv:2502.04217},
  year    = {2025},
}

@article{fischer2022nekrs,
  title   = {{NekRS}, a {GPU}-accelerated spectral element {Navier--Stokes} solver},
  author  = {Fischer, Paul and Kerkemeier, Stefan and Min, Misun and Lan, Yu-Hsiang and Phillips, Malachi and Rathnayake, Thilina and Merzari, Elia and Tomboulides, Ananias and Karakus, Ali and Chalmers, Noel and Warburton, Tim},
  journal = {Parallel Computing},
  volume  = {114},
  pages   = {102982},
  year    = {2022},
  doi     = {10.1016/j.parco.2022.102982},
}

@article{barwey2025mesh,
  title   = {Mesh-based super-resolution of fluid flows with multiscale graph neural networks},
  author  = {Barwey, Shivam and Pal, Pinaki and Patel, Saumil and Balin, Riccardo and Lusch, Bethany and Vishwanath, Venkatram and Maulik, Romit and Balakrishnan, Ramesh},
  journal = {Computer Methods in Applied Mechanics and Engineering},
  volume  = {443},
  pages   = {118072},
  year    = {2025},
  doi     = {10.1016/j.cma.2025.118072},
}

@article{hannachi2007eof,
  author  = {Hannachi, A. and Jolliffe, I. T. and Stephenson, D. B.},
  title   = {Empirical {O}rthogonal {F}unctions and {R}elated {T}echniques in {A}tmospheric {S}cience: {A} {R}eview},
  journal = {International Journal of Climatology},
  year    = {2007},
  volume  = {27},
  number  = {9},
  pages   = {1119--1152},
  doi     = {10.1002/joc.1499},
}

@article{kalnay1996ncep,
  title={The {NCEP/NCAR} 40-year reanalysis project},
  author={Kalnay, E. and Kanamitsu, M. and Kistler, R. and Collins, W. and Deaven, D. and Gandin, L. and Iredell, M. and Saha, S. and White, G. and Woollen, J. and Zhu, Y. and Chelliah, M. and Ebisuzaki, W. and Higgins, W. and Janowiak, J. and Mo, K. C. and Ropelewski, C. and Wang, J. and Leetmaa, A. and Reynolds, R. and Jenne, R. and Joseph, D.},
  journal={Bulletin of the American Meteorological Society},
  volume={77},
  number={3},
  pages={437--472},
  year={1996},
  publisher={American Meteorological Society}
}

@article{kilmer2021tensor,
  title={Tensor-tensor algebra for optimal representation and compression of multiway data},
  author={Kilmer, Misha E and Horesh, Lior and Avron, Haim and Newman, Elizabeth},
  journal={Proceedings of the National Academy of Sciences},
  volume={118},
  number={28},
  pages={e2015851118},
  year={2021},
  publisher={National Academy of Sciences}
}

@article{jiang2020pca,
  author  = {Jiang, Yujing and Cooley, Daniel and Wehner, Michael F.},
  title   = {Principal Component Analysis for Extremes and Application to {U.S.} Precipitation},
  journal = {Journal of Climate},
  year    = {2020},
  volume  = {33},
  number  = {15},
  doi     = {10.1175/JCLI-D-19-0413.1},
}

@book{
	Ballard_Kolda_2025,
	place={Cambridge},
	title={Tensor Decompositions for Data Science},
	publisher={Cambridge University Press},
	author={Ballard, Grey and Kolda, Tamara G.},
	year={2025}
}

@article{newman2025optimal,
author = {Newman, Elizabeth and Keegan, Katherine},
title = {Optimal Matrix-Mimetic Tensor Algebras via Variable Projection},
journal = {SIAM Journal on Matrix Analysis and Applications},
volume = {46},
number = {3},
pages = {1764-1790},
year = {2025},
doi = {10.1137/24M1702635},

URL = {

        https://doi.org/10.1137/24M1702635



},
eprint = {

        https://doi.org/10.1137/24M1702635



}
}

@article{keegan2025projected,
  author  = {Keegan, Katherine and Newman, Elizabeth},
  title   = {Projected Tensor-Tensor Products for Efficient Computation of Optimal Multiway Data Representations},
  journal = {Linear Algebra and its Applications},
  volume  = {729},
  pages   = {100--147},
  year    = {2025},
  doi     = {10.1016/j.laa.2025.09.018}
}

@article{kernfeld2015tensor,
  author  = {Kernfeld, Eric and Kilmer, Misha and Aeron, Shuchin},
  title   = {Tensor-Tensor Products with Invertible Linear Transforms},
  journal = {Linear Algebra and its Applications},
  volume  = {485},
  pages   = {545--570},
  year    = {2015},
  doi     = {10.1016/j.laa.2015.07.021}
}

@article{eckart1936approximation,
  title   = {The Approximation of One Matrix by Another of Lower Rank},
  author  = {Eckart, Carl and Young, Gale},
  journal = {Psychometrika},
  volume  = {1},
  number  = {3},
  pages   = {211--218},
  year    = {1936},
  doi     = {10.1007/BF02288367}
}

@article{mirsky1960symmetric,
  title   = {Symmetric Gauge Functions and Unitarily Invariant Norms},
  author  = {Mirsky, Leon},
  journal = {The Quarterly Journal of Mathematics},
  volume  = {11},
  number  = {1},
  pages   = {50--59},
  year    = {1960},
  doi     = {10.1093/qmath/11.1.50}
}

@article{hitchcock1927expression,
  title   = {The Expression of a Tensor or a Polyadic as a Sum of Products},
  author  = {Hitchcock, Frank L.},
  journal = {Journal of Mathematics and Physics},
  volume  = {6},
  number  = {1--4},
  pages   = {164--189},
  year    = {1927},
  doi     = {10.1002/sapm192761164}
}

@article{carroll1970analysis,
  title   = {Analysis of Individual Differences in Multidimensional Scaling via an {N}-way Generalization of ``{E}ckart-{Y}oung'' Decomposition},
  author  = {Carroll, J. Douglas and Chang, Jih-Jie},
  journal = {Psychometrika},
  volume  = {35},
  number  = {3},
  pages   = {283--319},
  year    = {1970},
  doi     = {10.1007/BF02310791}
}

@article{harshman1970foundations,
  title   = {Foundations of the {PARAFAC} Procedure: Models and Conditions for an ``Explanatory'' Multi-modal Factor Analysis},
  author  = {Harshman, Richard A.},
  journal = {UCLA Working Papers in Phonetics},
  volume  = {16},
  pages   = {1--84},
  year    = {1970}
}

@article{tucker1966some,
  title   = {Some Mathematical Notes on Three-mode Factor Analysis},
  author  = {Tucker, Ledyard R.},
  journal = {Psychometrika},
  volume  = {31},
  number  = {3},
  pages   = {279--311},
  year    = {1966},
  doi     = {10.1007/BF02289464}
}

@article{delathauwer2000multilinear,
  title   = {A Multilinear Singular Value Decomposition},
  author  = {De Lathauwer, Lieven and De Moor, Bart and Vandewalle, Joos},
  journal = {SIAM Journal on Matrix Analysis and Applications},
  volume  = {21},
  number  = {4},
  pages   = {1253--1278},
  year    = {2000},
  doi     = {10.1137/S0895479896305696}
}

@article{delathauwer2000bestrank,
  title   = {On the Best Rank-1 and Rank-($R_1, R_2, \dots, R_N$) Approximation of Higher-Order Tensors},
  author  = {De Lathauwer, Lieven and De Moor, Bart and Vandewalle, Joos},
  journal = {SIAM Journal on Matrix Analysis and Applications},
  volume  = {21},
  number  = {4},
  pages   = {1324--1342},
  year    = {2000},
  doi     = {10.1137/S0895479898346995}
}

@article{oseledets2011tensor,
  title   = {Tensor-Train Decomposition},
  author  = {Oseledets, Ivan V.},
  journal = {SIAM Journal on Scientific Computing},
  volume  = {33},
  number  = {5},
  pages   = {2295--2317},
  year    = {2011},
  doi     = {10.1137/090752286}
}

@article{phipps2019software,
  title   = {Software for Sparse Tensor Decomposition on Emerging Computing Architectures},
  author  = {Phipps, Eric T. and Kolda, Tamara G.},
  journal = {SIAM Journal on Scientific Computing},
  volume  = {41},
  number  = {3},
  pages   = {C269--C290},
  year    = {2019},
  doi     = {10.1137/18M1210691},
  note    = {GenTen: shared-memory/Kokkos parallel CP decomposition}
}

@inproceedings{smith2016medium,
  title     = {A Medium-Grained Algorithm for Distributed Sparse Tensor Factorization},
  author    = {Smith, Shaden and Karypis, George},
  booktitle = {2016 IEEE International Parallel and Distributed Processing Symposium (IPDPS)},
  pages     = {902--911},
  year      = {2016},
  doi       = {10.1109/IPDPS.2016.113},
  note      = {SPLATT: distributed-memory parallel CP decomposition}
}

@inproceedings{austin2016parallel,
  title     = {Parallel Tensor Compression for Large-Scale Scientific Data},
  author    = {Austin, Woody and Ballard, Grey and Kolda, Tamara G.},
  booktitle = {2016 IEEE International Parallel and Distributed Processing Symposium (IPDPS)},
  pages     = {912--922},
  year      = {2016},
  doi       = {10.1109/IPDPS.2016.67}
}

@article{ballard2020tuckermpi,
  title   = {{TuckerMPI}: A Parallel {C}++/{MPI} Software Package for Large-Scale Data Compression via the {Tucker} Tensor Decomposition},
  author  = {Ballard, Grey and Klinvex, Alicia and Kolda, Tamara G.},
  journal = {ACM Transactions on Mathematical Software},
  volume  = {46},
  number  = {2},
  pages   = {1--31},
  year    = {2020},
  doi     = {10.1145/3378445}
}

@Article{BKK20,
  author      = {Ballard, Grey and Klinvex, Alicia and Kolda, Tamara G.},
  title       = {{TuckerMPI}: A Parallel {C++/MPI} Software Package for Large-Scale Data Compression via the Tucker Tensor Decomposition},
  journal     = {ACM Transactions on Mathematical Software},
  year        = {2020},
  volume      = {46},
  number      = {2},
  month       = jun,
  address     = {New York, NY, USA},
  articleno   = {13},
  doi         = {10.1145/3378445},
  issn        = {0098-3500},
  issue_date  = {June 2020},
  numpages    = {31},
  publisher   = {ACM},
  techversion = {BKK19-TR},
  url         = {https://dl.acm.org/doi/10.1145/3378445},
}

@Article{ABB22,
  author      = {Al Daas, Hussam and Ballard, Grey and Benner, Peter},
  journal     = {SIAM Journal on Scientific Computing},
  title       = {Parallel Algorithms for Tensor Train Arithmetic},
  year        = {2022},
  number      = {1},
  pages       = {C25-C53},
  volume      = {44},
  doi         = {10.1137/20M1387158},
  techversion = {ABB20-TR},
  url         = {https://doi.org/10.1137/20M1387158},
}

@InProceedings{HL+21,
  author    = {Helal, Ahmed E. and Laukemann, Jan and Checconi, Fabio and Tithi, Jesmin Jahan and Ranadive, Teresa and Petrini, Fabrizio and Choi, Jeewhan},
  booktitle = {Proceedings of the 35th ACM International Conference on Supercomputing},
  title     = {{ALTO}: adaptive linearized storage of sparse tensors},
  year      = {2021},
  address   = {New York, NY, USA},
  pages     = {404?416},
  publisher = {Association for Computing Machinery},
  series    = {ICS '21},
  doi       = {10.1145/3447818.3461703},
  isbn      = {9781450383356},
  numpages  = {13},
  url       = {https://doi.org/10.1145/3447818.3461703},
}

@Article{EH+21,
  author      = {Eswar, Srinivas and Hayashi, Koby and Ballard, Grey and Kannan, Ramakrishnan and Matheson, Michael A. and Park, Haesun},
  title       = {{PLANC}: Parallel Low-Rank Approximation with Nonnegativity Constraints},
  journal     = {ACM Transactions on Mathematical Software},
  year        = {2021},
  volume      = {47},
  number      = {3},
  month       = jun,
  address     = {New York, NY, USA},
  articleno   = {20},
  doi         = {10.1145/3432185},
  issn        = {0098-3500},
  issue_date  = {June 2021},
  numpages    = {37},
  publisher   = {ACM},
  techversion = {EH+19-TR},
  url         = {https://doi.org/10.1145/3432185},
}

@inproceedings{WKP25,
author = {Wijeratne, Sasindu and Kannan, Rajgopal and Prasanna, Viktor},
title = {AMPED: Accelerating MTTKRP for Billion-Scale Sparse Tensor Decomposition on Multiple GPUs},
year = {2025},
isbn = {9798400720741},
publisher = {Association for Computing Machinery},
address = {New York, NY, USA},
url = {https://doi.org/10.1145/3754598.3754651},
doi = {10.1145/3754598.3754651},
booktitle = {Proceedings of the 54th International Conference on Parallel Processing},
pages = {208–217},
numpages = {10},
location = {
},
series = {ICPP '25}
}

@InProceedings{KU16,
  author    = {O. Kaya and B. U\c{c}ar},
  title     = {High Performance Parallel Algorithms for the {T}ucker Decomposition of Sparse Tensors},
  booktitle = {45th International Conference on Parallel Processing (ICPP '16)},
  year      = {2016},
  pages     = {103-112},
  doi       = {http://dx.doi.org/10.1109/ICPP.2016.19},
}

@article{RTB22,
author = {R\"{o}hrig-Z\"{o}llner, Melven and Thies, Jonas and Basermann, Achim},
title = {Performance of the Low-Rank TT-SVD for Large Dense Tensors on Modern MultiCore CPUs},
journal = {SIAM Journal on Scientific Computing},
volume = {44},
number = {4},
pages = {C287-C309},
year = {2022},
doi = {10.1137/21M1395545},
URL = {https://doi.org/10.1137/21M1395545}
}

@article{dghklty14,
    title     = {Accelerating Numerical Dense Linear Algebra Calculations with GPUs},
    author    = {Jack Dongarra and Mark Gates and Azzam Haidar and Jakub Kurzak and 
                 Piotr Luszczek and Stanimire Tomov and Ichitaro Yamazaki},
    journal   = {Numerical Computations with GPUs},
    pages     = {1-26},
    year      = {2014},
    publisher = {Springer}
}

@manual{MKL,
  title        = {Developer Reference for Intel® oneAPI Math Kernel Library},
  author       = {{Intel Corporation}},
  year         = {2026},
  url          = {https://www.intel.com/content/www/us/en/developer/tools/oneapi/onemkl-documentation.html},
}

@article{AC+21,
author = {Abdelfattah, Ahmad and Costa, Timothy and Dongarra, Jack and Gates, Mark and Haidar, Azzam and Hammarling, Sven and Higham, Nicholas J. and Kurzak, Jakub and Luszczek, Piotr and Tomov, Stanimire and Zounon, Mawussi},
title = {A Set of Batched Basic Linear Algebra Subprograms and LAPACK Routines},
year = {2021},
issue_date = {September 2021},
publisher = {Association for Computing Machinery},
address = {New York, NY, USA},
volume = {47},
number = {3},
issn = {0098-3500},
url = {https://doi.org/10.1145/3431921},
doi = {10.1145/3431921},
journal = {ACM Trans. Math. Softw.},
month = jun,
articleno = {21},
numpages = {23}
}

@article{kolda2009tensor,
  title   = {Tensor Decompositions and Applications},
  author  = {Kolda, Tamara G. and Bader, Brett W.},
  journal = {SIAM Review},
  volume  = {51},
  number  = {3},
  pages   = {455--500},
  year    = {2009},
  doi     = {10.1137/07070111X}
}

@misc{KVX21,
      title={A Tensor SVD-based Classification Algorithm Applied to fMRI Data}, 
      author={Katherine Keegan and Tanvi Vishwanath and Yihua Xu},
      year={2021},
      eprint={2111.00587},
      archivePrefix={arXiv},
      primaryClass={cs.LG},
      url={https://arxiv.org/abs/2111.00587}, 
}

\end{document}